**Twenty Constructionist Things to Do with Artificial Intelligence and Machine Learning**

Yasmin Kafai & Luis Morales-Navarro, University of Pennsylvania

**ABSTRACT**
In this paper, we build on the 1971 memo "Twenty Things to Do With a Computer" by Seymour Papert and Cynthia Solomon and propose twenty constructionist things to do with artificial intelligence and machine learning. Several proposals build on ideas developed in the original memo while others are new and address topics in science, mathematics, and the arts. In reviewing the big themes, we notice a renewed interest in children's engagement not just for technical proficiency but also to cultivate a deeper understanding of their own cognitive processes. Furthermore, the ideas stress the importance of designing personally relevant AI/ML applications, moving beyond isolated models and off-the-shelf datasets disconnected from their interests. We also acknowledge the social aspects of data production involved in making AI/ML applications. Finally, we highlight the critical dimensions necessary to address potential harmful algorithmic biases and consequences of AI/ML applications.

This paper builds on a venerable tradition in the constructionist community, starting with "Twenty Things to Do With a Computer" that Seymour Papert and Cynthia Solomon published in 1971 as part of the MIT AI Lab Memos series. The memo presented a collection of different ideas and applications in the arts and sciences, and provided diverse entry points—some easy, others more difficult—into computing for children, youth, and teachers. The idea was to challenge conventions about what one could do with computers while also gaining deeper insights into one's own thinking and computing. This tradition has been picked up when imagining things to do with programmable bricks (Resnick et al., 1996), with Scratch (Hideki, 2017), or with living materials in biology (Kafai & Walker, 2020). A recent book-length edition of essays (Stager, 2021) reviews these ideas fifty years later. Underlying all of these collections is a common thread of making things with computing rather than just observing or using them—an effort which we are continuing with a first attempt at twenty constructionist things to do with artificial intelligence and machine learning.

In the last decade there has been an exponential growth in artificial intelligence and machine learning (AI/ML) applications, moving out of the lab into the world, impacting everyday lives. There is now a growing recognition that all students and teachers need to be prepared for understanding and using AI/ML applications (Long & Magerko, 2020; Touretzky et al., 2019). However, to date most efforts promoting artificial intelligence in education have focused on what Eisenberg and colleagues (2017) called artificial-teachers, artificial-tutors, or artificial co-learners, centering learners as recipients of instruction from AI/ML agents or collaborators with AI/ML agents (Ouyang & Jiao, 2021) with much less attention paid to learners creating applications and designing culturally relevant and critically responsive AI/ML projects. Papert and Solomon (1971) observed that while most approaches to computing in education envisioned that "the transaction between a computer and the kid will be some kind of 'conversation' or 'questions and answer' in words or numbers," (p.1) there are many other ways in which children could interact and create with computers. In *Mindstorms: Children, Computers and Powerful Ideas* (1980), Papert more clearly delineated two different visions for how children could interact and learn with computing, the instructionist one focusing on "the computer being used to program the child" (p. 5) and the constructionist one in which "the child programs the computer and, in doing so, both acquires a sense of mastery over a piece of the most modern and powerful technology and establishes an intimate contact with some of the deepest ideas from science, from mathematics, and from the art of intellectual model building" (p. 5). No matter which vision one adopts, it is the constructionist vision, in which the child learns to create with computing, which will be the focus of this paper and how we can think about learning AI/ML.

Constructionism, the theory and pedagogy, itself was born in the context of artificial intelligence research in the 1970s. As Kahn and Winters (2021) and Solomon and colleagues (2020) discuss issues of education and artificial intelligence were part of the early development of LOGO. But it is only recently that constructionist researchers have turned their attention again to the design of learning environments for children to learn and create with AI/ML (Kahn & Winters, 2021). Kahn (1977) initially outlined three different roles that artificial intelligence could play in education, one of them was having students use AI programming tools in their projects, a second was having students create artifacts by interacting with AI, and a third involved the use of computational theories of intelligence and learning in the design of learning activities. These roles were later expanded to include five approaches for constructionist AI and education (Kahn & Winters, 2020), among them children learning AI/ML as a school subject, children engaging with AI/ML to reflect on their own learning and problem solving, applying AI/ML to solve problems in other subject domains, learning by creating AI/ML open-ended projects, and engaging with the social implications of AI/ML through building and critiquing AI/ML models embedded in applications. While contemporary AI/ML technologies differ from those of early AI research by relying on ML and large data sets instead of symbolic AI, those early questions concerning how children think about their own thinking with computing and AI/ML remain significant. Additionally, it is essential to support young people in developing AI/ML-powered applications so that they are equipped to participate in computing fully and critically (DiPaola et al., 2022; Kafai & Proctor, 2022).

In the following sections, we present a list of 20 constructionist things to do with AI/ML that embrace a range of different applications. Many of these things involve using available tools while others might require making your own. Already the original paper "Twenty Things To Do With A Computer" (1971) suggested several activities that could become AI/ML-powered projects such as programming robots with sensors (#1), composing music (#7 and #4), and writing poetry (#5) and have been included here. We also draw on work by Eisenberg and colleagues' (2017) revaluations of minds and machines, Kahn and Winters (2021) outlining what they called "half-baked AI projects, Ojeda-Ramirez's and colleagues' metacognitive reflections for ideas, and many others on how learners can engage creatively with AI. In addition, several ideas were contributed by conference participants at Fablearn/Constructionism 2023.

# 1 Smart Puppets

In the original memo, Papert and Solomon (#19, 1971) propose making puppets and using motors to control them. With AI/ML we can extend this idea to create more interactive puppets that can recognize words and speech or movements. Older children and youth could have puppets recognize speech or words (from using simple classifiers to recognize keywords to more complex voice recognition libraries) and respond to them accordingly (with preset responses—à la Eliza— or using synthetic text and text to speech libraries). Younger children, as Tseng and colleagues (2021) demonstrate, could use plushies to tell stories and build classifiers of plushie movements to trigger sounds and voice recordings.

# 2 Face Filter

Young people interact with AI/ML powered face filters when using social media (SnapChat, TikTok) with their friends. Creating filters can involve them in learning about AI/ML in two ways: exploring how to recognize facial features and body parts and using generative models to create images to transform what is captured or superposed over the camera feed. Making face filters can be a powerful way for learners to explore how something that they interact with everyday works, to inquire into why some filters work better for some people, and to engage with ideas of algorithmic justice in a personally relevant way.

#3 Weird Recipes

Tseng and colleagues (2023) show how youth can use classifiers to build an application that provides recipe recommendations. Using libraries such as nanoGPT, could enable learners to play with language models, modify training data sets, and change parameters to create recipes. Learners could also cook the recipes and try their results. At Fablearn/Constructionism 2023, Jaymes Dec suggested creating a further iteration, creating an application that provides weird recipes mixing ingredients and cooking steps.

# 4 Dance Game

Dance in AI/ML education has been growing over the past few years with projects that have learners partner to dance with AI agents (Long et al., 2020) and have learners create choreographies and animations (Castro et al., 2022; Jordan et al., 2021. Building on these efforts, children could design their own dance games similar to Just Dance, where they train models to recognize players' moves (sequences of poses) using either pose recognition or data from wearable sensors and compare these to predetermined choreographies or where based on a set of basic moves the system generates possible choreographies that match music rhythms. In building their games, children could incorporate their own music and dance taste, create personalized animations and test the games with their peers. Making dance games should involve peer-testing of applications so that learners can iteratively identify edge and failure cases and improve their projects (Morales-Navarro et al., 2024).

# 5 Poetry Generator

Papert and Solomon (#15, 1971) explored how creating sentence generators may support learners to better understand language structures and computing. With current off-the-shelf AI/ML libraries, learners can generate poetry while exploring more sophisticated natural language processing techniques. From simple Markov chains to training models using nanoGPT, youth can create synthetic text including poetry, short stories, and music lyrics. Creating synthetic text can also provide opportunities to inquire and discuss critical issues related to disinformation, copyright, creativity, the limitations of synthetic text, and the amount of energy that language models may consume.

# 6 Sports Training App

Similarly, to #4, there's a lot of promise for young people to learn about AI/ML while engaging with sports (Jordan et al., 2021; Kumar & Worsley, 2023; Zimmermann-Niefield; 2019). Here youth could create applications that recognize athletic moves and support them in improving their form (e.g., when pitching a ball). Other applications could imitate commercial fitness apps, supporting youth to create models with data generated with their own bodies (e.g., heartbeat, body temperature, speed). These types of applications could also serve as opportunities to discuss issues of data privacy and ethics.

# 7 Music Generator

Papert & Solomon (# 11, 1971) suggested creating a music box with programmed tunes. Later in #12, they present a project that creates random songs. Today, children could build data sets of tunes and use them to train a model that generates music, these "compositions" could follow the traditions of aleatoric and minimalist music or even explore noise music. Here, like in #2 learners could engage in discussions about creativity in humans and machines and copyright.

# 8 Artificial Creatures and Systems for Natural Habitats

Building on advances in embodied cognition, Eisenberg and colleagues propose that children could "create free-standing simple creatures whose job is to be placed in a terrarium or aquarium alongside living creatures and to interact with their (biological) environment" (Eisenberg et al. 2017 p.3) to reflect on the embodied nature of their own thinking. Such projects would involve building AI/ML models for the systems to recognize environmental behaviors and react to them. This would also be a rich space for learners to think about technology not only as embodied and socio-technical but also as being in relationship with ecological systems. "Children might create birdhouses (to be placed in natural settings) that offer food and shelter preferentially to certain species of birds, or only under specific conditions of temperature or illumination; studying how birds respond to the artifact over time might be an introduction both to ontogenetic learning in animals and (over still longer periods) to the notion of evolutionary constraints." (Eisenberg et al.

2017 p.3) Such projects would also involve thinking about ethics and the environmental impact of children's creations.

# # 9 Game with a Game Player

Making games has always been a rich context for engaging with computing in constructionist ways. With AI/ML children cannot only create games to play with each other (#5 Papert & Solomon, 1971), but also train models that when incorporated into the games can enable them and their peers to play against the machine. Examples include creating, training, and then using neural networks to play TicTacToe as Kahn and Winter (2021) suggested in half-baked AI projects or by designing and playing a Rock-Paper-Scissors game with an application that recognizes speech and hand gestures (Kahn & Winter, 2018). These activities may support learners to think about how they think when playing games as they design and build models they can play against.

# # 10 Explain Yourself

In #18, Solomon and Papert (1971) invite learners to explain themselves asking "How good of a model could you make of a person?." A similar thing could be done with AI/ML, inviting learners to think about how the qualities of synthetic text differ from those of text written by humans, how good of a co-learner an AI/ML system can be, or how good of a player an AI agent can be. The explain yourself activity can serve as a space for learners to think about the differences between human and machine learning, to explore the limitations of both humans and machines and to think about their own thinking.

# # 11 Drawing Generator

Learners could create drawing generators that make images in comic or anime styles. Here they could use existing public image large datasets that they can modify and or personalize. While creating an image generator (using off-the-shelf libraries) learners could also reflect on the environmental impact of generating synthetic images (see Luccioni et al., 2023) and what are the ethics of this in the context of climate change. Similarly, this activity could also involve learners in discusions related to how synthetic images may contribute to misinformation.

# # 12 Adaptable Interactive Stories

At Fablearn/Constructionism 2023, Rita Freudenberg suggested that children could "create interactive stories [or games], where the storyline not just depends on a deliberate action by the player, but the presented story options change based on previous decisions". Rita explained learners would have to decide on what factors would influence the development of the narrative and how transparent these are to users. Such a project could involve children curating datasets together and training models to classify user inputs. As Ken Kahn proposed at Fablearn/Constructionism 2023 learners may also use generative models to create images to illustrate the interactive stories. While doing this, learners can explore questions about creativity and AI/ML, share stories with their peers, and evaluate each others' stories.

# # 13 Artificial Tutors for Peers

Similar to #9, learners could create a system that interacts with junior students. Fablearn/Constructionism 2023 participant Daniel Agostini argued that this could support learners to "learn by teaching" an AI/ML system. We extend Agostini's idea within a larger constructionist tradition of children and youth creating software for their younger peers (Harel, 1990). With AI/ML learners could create tutors for younger students. Here, learners could use ethical matrices to identify the values that people with different roles (designers, students, teachers, parents) may have with regards to the co-learners (DiPaola et al., 2020).

# #14 Modeling Climate and Carbon Emissions

Students could create models using publicly available climate data to recognize patterns and make predictions. This suggestion by an anonymous participant at Fablearn/Constructionism 2023, highlights how such an activity can engage learners in thinking about AI/ML in a highly relevant area of concern that is socio-technical and involves thinking about complexity. These kinds of activities could build on the rich history of constructionist research on scientific modeling and complex systems (Wilensky & Rand, 2015).

# # 15 Artificial Co-Learners

Eisenberg and colleagues also propose having learners create artificial co-learners. This, they describe, would involve having youth "design models of students (rather than teachers) that would accompany children in learning new material and articulate (perhaps simplistic, but concrete) strategies for thinking about the material. Such models might be customizable or programmable for students–that is, students might be able to create (in high-level form) particular rules for learning to see how those rules play out in the behavior of their artificial colleague" (Eisenberg et al. 2017 p.3). While doing this, learners should be encouraged to test their co-learners with their peers.

# #16 Role-Playing Games

Students could create games based on text inputs. We build on this idea of Fablearn/Constructionism 2023 participant Ken Kahn by proposing that learners could create role-playing games, for instance building datasets of role-playing game scripts (such as Dungeon and Dragons campaigns), training models (see #1) and then generating new games. The same could be done to generate fan fiction. This kind of activity could support learners in discussing how the data is produced in social contexts, think about the ethics of data usage, and the implications of using AI/ML in creative tasks.

# 17 Personal Assistants

During Fablearn/Constructionism 2023 Nicolás Acosta and Glenn Boustead suggested that learners could build personal assistants. Acosta and Boustead explained this could involve creating a voice assistant to control music, lights, and engage people in wellness activities (e.g., meditation, reflection). While building a personal assistant, Bousted suggested, learners could reflect on how AI/ML tools shape our everyday activities and how simple tasks such as scheduling involve ethical decisions. They say "it is less about the results [of the personal assistant] and more about the ethical nature of the process." Van Brummelen's (2022) work can serve as a foundation to further design tools and activities to support learners in creating personal assistants.

#18 Tamagotchi

Learners could create virtual pets that can speak and listen, imagining for example that someone might create a "tamagotchi with sass." In this proposal developed by Fablearn/Constructionism 2023 participant Jaymes Dec, students could for instance use off-the-shelf text to speech libraries, use language models (perhaps even train their own simple one like in #2), and exchange tamagotchi with their peers.

# 19 Workout App

During Fablearn/Constructionism 2023, Deborah Fields proposed that learners could create applications that provide users with personalized workout plans. Here learners could create simple models trained on workout plans apps could take into consideration user preferences and output a personalized workout plan. This type of project could also engage learners in thinking about the ethical implications of creating such an app with regards to how it may affect the health of the users.

#20 Create a constructionist activities generator

Create an application that uses AI/ML to generate more constructionist activities.

Each of the proposed twenty things to do provides compelling and valuable directions for engaging learners with AI/ML in constructionist ways, yet some aspects of learning AI/ML — metacognition, artifacts, collaboration, and ethics—deserve further consideration.

To start, having children engage with AI/ML to reflect on their own learning and problem solving was already part of initial constructionist efforts but then lost traction as economic, social, and ethical considerations moved to the fore (Eisenberg, Hsi & Oh, 2017; Ojeda-Ramirez et al., 2023). Eisenberg and colleagues (2017) remind us that considering the computer as a "psychological machine" influences how we think about our own thinking (Turkle, 1984) and outline several interesting proposals for computing and AI/ML to engage children in exploring their own intuitions about thinking. Likewise, Ojeda-Ramirez and colleagues (2023) position the value that

AI/ML can bring in helping learners reflect on their own thinking and learning as another rationale for teaching AI. They propose different approaches to AI/ML that can serve as metacognitive tools taking advantage of, for example, off-the-shelf large language model applications. A key theme here is to think about AI as a form of microworld that lets learners inquire about cognitive processes in more concrete ways. Yet, it is equally important to ensure that children have opportunities to understand and investigate that humans and machines "learn" *differently* and that "learning" in AI/ML systems has a very different meaning than in humans. This may be productive for learners to think about the meaning of *understanding* in both humans and machines (Mitchel & Krakauer, 2023), to recognize how their own thinking differs from how machines process data (Yiu, Kosoy, & Gopnik, 2023) and to explore the limitations of the computational metaphors of cognition (Marcus, 2014). While these inquiries might be difficult to accomplish because of the opacity of today's AI/ML models they are worthwhile investigations nonetheless. As Eisenberg and colleagues (2017) stated succinctly:

> "The discussions along these lines in *Mindstorms* were not uniformly triumphalist about computational models of mind. There was at least a suggestion in the book that programming could give children a means of reflecting on the distinct limitations both of humans and machines. Children were not (it appears) intended reflexively to identify their thinking with computational processes, but rather to use the computer as an imperfect mirror, a loose and intriguing model against which to compare their own minds and selves." (p. 1)

A second aspect builds on the idea of providing opportunities for youths to create personally relevant projects or applications that incorporate AI/ML. All the examples in the original 1971 memo by Papert and Solomon involve learners creating something, be it a game, a theater play or a poem to learn about powerful computing ideas rather than focusing on computing concepts. Later Papert (1991) captured this defining aspect of constructionist learning stating that learning "happens especially felicitously in a context where the learner is consciously engaged in constructing a public entity, whether it's a sandcastle on the beach or a theory of the universe" (p. 1). This shift from learning isolated computing concepts in a vacuum to designing personally relevant applications has in present days become a mainstay in most K-12 computing education initiatives (Kafai & Burke, 2014, Sentance & Waite, 2021). Yet, often AI/ML education efforts focus on having learners create models in a vacuum or use off-the-shelf datasets that may not relate to learners' interests. Zimmermann-Niefield and colleagues (2019), for example, have researched the opportunities and challenges that children encounter when they incorporate model building into traditional constructionist computing projects. Here it is key to think of learning about AI/ML not as an end but as a tool for youths to express themselves and participate in the world.

A third aspect is the increased attention being paid to ethics addressing potential algorithmic harmful biases of AI/ML applications (Benjamin, 2021; Birhane, 2021). Recent work has begun to examine how youth understand algorithmic justice, what funds of knowledge are used by children to judge algorithmic systems, and how children's sensemaking can be scaffolded (Salac, Landesman, Druga, & Ko, 2023; Coenrad, 2022). While previous studies with youth have often centered on high stakes contexts such as policing and surveillance (eg., Vakil & McKinney de Royston, 2022; Walker et al., 2022), research also now focuses on popular photo and video filters in social media applications such as TikTok or Snapchat, streaming services, voice assistants, and writing assistants (see Festerling & Siraj, 2020) which have been present in youth everyday lives for more than a decade. A key theme here is how critical inquiry can become part of the design process and understanding of algorithmic justice (Ali et al, 2022). Beyond harmful biases, we also need to expand critical inquiries into the ecological impact of current use of AI/ML due to their extensive energy consumption (Luccioni, 2023; Crawford, 2021). Other important ethical issues that should be explored with youth include the role of AI/ML in producing misinformation (Zhou et al., 2023), the ethics of data usage, and the often ignored essential human labor involved in data production (Amrute, 2019; Crawford, 2021).

A fourth aspect addresses the social dimensions of AI/ML applications. For one, Papert (1991) stated that creating publicly shareable artifacts or applications is key to learning. Creating applications is not just for personal fulfillment but also has social purposes. Furthermore, the social nature of data production in many AI/ML applications is often ignored. In fact, producing, collecting and testing data are not passive endeavors but inherently social processes that involve the selection of tools that produce data (including the knowledge and values embedded in them), and the human purposes for which data are created. For these reasons, Hardy, Dixon and Shi (2020) argued for more explicit student agency in data production so that learners can develop an understanding of data as produced "through interaction between instruments and the material world, using technology that embody knowledge and values, in pursuit of human goals and purposes" (p. 123). While their focus was on situating data production within the context of science inquiry, it also applies to data produced, labeled, and used in AI/ML applications. Here peers can also play an important part supporting other learners to take perspective and consider the limitations and implications of their designs, through testing (Morales-Navarro, Shah, & Kafai, 2023). An expansion would be the introduction of peer auditing, where not the designers themselves but others would investigate functional and bias of AI/ML applications (Metaxa et al., 2021).

Finally, while we discussed each aspect individually, in many of the proposed AI/ML activities they come together, fruitfully informing each other. As illustrated in #6, discussions about collecting data about sports activities and performance do not need to be limited to the technicalities but can also provide rich context for addressing issues of privacy and ethics. Likewise, #10 raises central issues about the limitations of machine thinking, when compared to human thinking. All the examples (and there are many more to think of) provide a rich tapestry of possibilities of how learners can engage creatively with artificial intelligence and machine learning, rather than being on the receiving end of technology, and promote a deeper understanding about oneself, others, and the world while also furthering knowledge about the potentials and limitations of technology.

**Acknowledgments**
This writing of this paper was supported by National Science Foundation grant #2333469. Any opinions, findings, and conclusions or recommendations expressed in this paper are those of the authors and do not necessarily reflect the views of NSF or the University of Pennsylvania. We thank Nathan Holbert for his comments on an earlier draft of the paper. We also thank the members of the Constructionism community for their various suggestions of what to include in the list.